\documentclass{nature}

\usepackage[english]{babel}
\usepackage[T1]{fontenc}
\usepackage{csquotes}
\usepackage[fleqn]{amsmath}
\usepackage{amsmath}
\usepackage{pdfpages}
\usepackage{wasysym}
\usepackage{amsmath}
\usepackage[version=4]{mhchem}
\usepackage{array}
\usepackage{graphicx}

\usepackage[labelfont=bf]{caption}
\makeatletter
\let\saved@includegraphics\includegraphics
\AtBeginDocument{\let\includegraphics\saved@includegraphics}
\renewenvironment*{figure}{\@float{figure}}{\end@float}
\makeatother

\title{Autonomous discovery of battery electrolytes with robotic experimentation and machine-learning}

\begin{document}

\author{Adarsh Dave$^{1,4}$, Jared Mitchell$^{2}$, Kirthevasan Kandasamy$^{3}$, Sven Burke$^{2}$, Biswajit Paria$^{3}$, Barnabas Poczos$^{3}$, Jay Whitacre$^{2,4,*}$, Venkatasubramanian Viswanathan$^{1,4,*}$}

\maketitle

\begin{affiliations}
 \item Department of Mechanical Engineering, Carnegie Mellon University, Pittsburgh, Pennsylvania, 15213, USA.
 \item Department of Materials Science and Engineering, Carnegie Mellon University, Pittsburgh, Pennsylvania, 15213, USA.
  \item Department of Machine Learning, Carnegie Mellon University, Pittsburgh, Pennsylvania, 15213, USA.
 \item Wilton E. Scott Institute for Energy Innovation, Carnegie Mellon University, 
 Pittsburgh,Pennsylvania, 15213, USA.
 \item[*] Corresponding Authors: whitacre@andrew.cmu.edu (J. W.), venkvis@cmu.edu (V. V.)
\end{affiliations}

\maketitle
\begin{abstract}
Innovations in batteries take years to formulate and commercialize, requiring extensive experimentation during the design and optimization phases. We approached the design and selection of a battery electrolyte through a black-box optimization algorithm directly integrated into a robotic test-stand. We report here the discovery of a novel battery electrolyte by this experiment completely guided by the machine-learning software without human intervention. Motivated by the recent trend toward super-concentrated aqueous electrolytes\cite{suo_water--salt_2015,zheng_understanding_2018,yokoyama_origin_2018} for high-performance batteries, we utilize Dragonfly - a Bayesian machine-learning software package\cite{kandasamy_tuning_2019} - to search mixtures of commonly used lithium and sodium salts for super-concentrated aqueous electrolytes with wide electrochemical stability windows. Dragonfly autonomously managed the robotic test-stand, recommending electrolyte designs to test and receiving experimental feedback in real time. In 40 hours of continuous experimentation over a four-dimensional design space with millions of potential candidates, Dragonfly discovered a novel, mixed-anion aqueous sodium electrolyte with a wider electrochemical stability window than state-of-the-art sodium electrolyte. A human-guided design process may have missed this optimal electrolyte. This result demonstrates the possibility of integrating robotics with machine-learning to rapidly and autonomously discover novel battery materials.
\end{abstract}

Energy-dense and safe batteries are crucial for electrification of transportation\cite{Deng:2018aa} and aviation\cite{Viswanathan2019}. But even incremental improvements to battery materials can take years to deliver, involving many rounds of iterative testing to optimize numerous material parameters to achieve multiple objectives\cite{tabor_accelerating_2018}. The battery design problem is fundamentally a complex function that takes battery formulation as input and returns performance measurements as output. Machine-learning methods can be used to optimize these black-box functions, even under multiple objectives\cite{pmlr-v37-kandasamy15,pmlr-v97-kandasamy19a,PariaKP19,kandasamy_tuning_2019,HernandezLobato16}. Machine-learning coupled with automated evaluation - whether via robotic experimentation or automated simulation workflows, able to immediately act on the model's recommendations - can ``close the loop'' and enable inverse material design\cite{kusne_fly_2014,zunger_inverse_2018,granda_controlling_2018, bhowmik_perspective_2019,bai_accelerated_2019,sun_accelerated_2019}. Bayesian optimization in particular has proven effective in solving chemical design problems over minimal experimental iterations, with successful examples in fields such as carbon nanotube\cite{nikolaev_discovery_2014} and polymer fiber synthesis\cite{li_rapid_2017}, metamaterial design\cite{bessa2019}, and organic photovoltaic devices\cite{macleod_self-driving_2019}. While similar approaches have been attempted in several fields of study, to our knowledge, this is the first attempt to apply this framework to the design of functional materials in electrochemistry. 

\begin{figure}
    \centering
    \includegraphics[width=\textwidth]{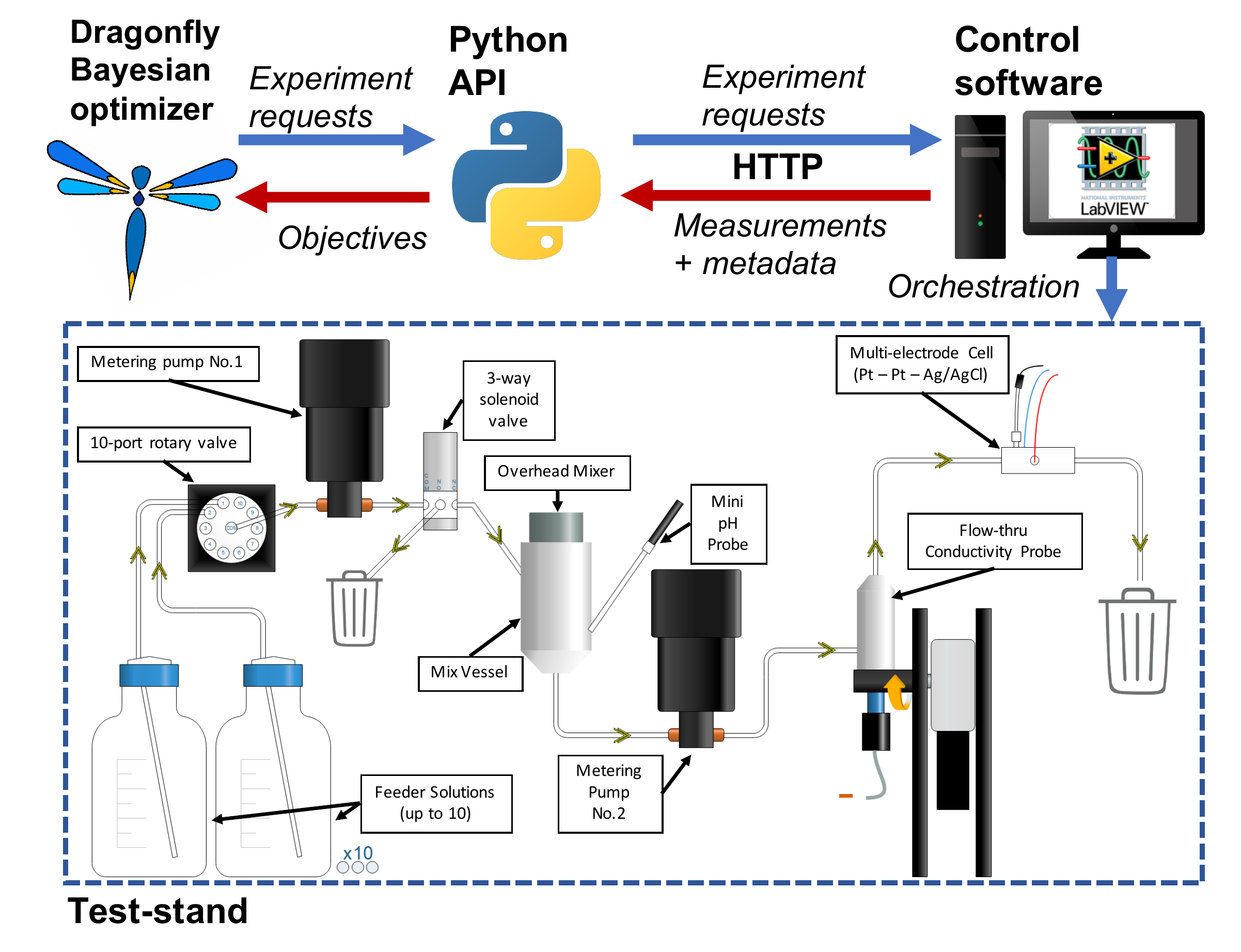}
    \caption{Schematic figure of Otto system illustrating test-stand and software architecture. The test-stand is capable of mixing arbitrary mixtures of chosen feeder solutions (salts dissolved into water near saturation) and measuring pH, ionic conductivity, and electrochemical properties in a symmetric platinum electrode cell. The control software takes in an experiment request (i.e. an electrolyte mixture to be characterized) and returns measurements and metadata (e.g. temperature) over HTTP to a Python API. Dragonfly - the Bayesian optimization software utilized in this paper - plugs into the Python API for requesting experiments and receiving feedback. A more detailed version on test-stand components and logic is available as Extended Data Figure 8.}
    \label{fig:schematic_lite}
\end{figure}

We have built a robotic platform for characterizing battery electrolytes\cite{whitacre_autonomous_2019,dave_benchmarking_2020}, shown in schematic in Figure 1; we allow this platform to run autonomously, guided by a machine-learning optimizer that plans each experimental iteration sequentially based on real-time experimental feedback. In this work, we demonstrate the use of this platform to optimize for a single objective - the electrolyte's electrochemical stability window  - in both  lithium-ion and sodium-ion aqueous electrolytes. We report a novel, high-performing dual-anion sodium electrolyte discovered by this platform over just 40 hours of continuous experimentation given four common sodium salts to choose from. The blended electrolyte is measured to have a wider electrochemical stability window than state-of-the-art sodium electrolyte\cite{lee_toward_2019}, suggesting a longer calendar life and improved high-rate capability over the state-of-the-art system. This result illustrates the promise of using machine-learning coupled to robotic experiments to rapidly optimize material designs and discover designs that human experimenters may miss.

The non-aqueous electrolytes commonly used in modern batteries are highly flammable, and present significant safety hazards and manufacturing costs relating to safety, storage, and management\cite{luo_raising_2010,suo_water--salt_2015}. Aqueous electrolytes are an attractive alternative. They are safer, lower-cost, and more conductive that non-aqueous counterparts\cite{li_towards_2013,whitacre_polyionic_2015}. High conductivity particularly suits large-format batteries that may be used in the electrical grid to smooth out the intermittent generation of power from renewable sources\cite{wu_relating_2015}.

Aqueous electrolytes have a narrow electrochemical stability window, limited by the hydrogen evolution reaction (HER) at low electrochemical potentials and the oxygen evolution reaction (OER) at high potentials\cite{li_rechargeable_1994}. These parasitic reactions preclude the use of the modern, high-voltage electrode couples that enable the high energy density of non-aqueous batteries\cite{luo_raising_2010}, and lead to poor cycling capability, calendar life, and diminished high-rate performance\cite{luo_aqueous_2007}. A recent trend in aqueous electrolyte design uses very high salt concentrations to suppress these reactions, either by deposition of a kinetically passivating electrode film (generally via anionic reduction) or by modifying interfacial hydration structures to achieve similar effects. These ``water-in-salt'' electrolytes have been shown to expand the electrochemical stability window from less than 2V for a standard aqueous electrolyte up to 3V, with effects demonstrated for a wide variety of salts \cite{suo_water--salt_2015,zheng_understanding_2018,yokoyama_origin_2018,lee_toward_2019}. Water concentration alone has been shown to have asymmetric influence on electrolyte resistance to HER and OER\cite{yokoyama_origin_2018}. Blending salts in electrolytes can positively impact performance, opening the possibility for mixed-anion electrolytes to have an optimal electrochemical window\cite{weber_long_2019,suo_advanced_2016}. This creates a design problem where the electrochemical stability window of an aqueous electrolyte could be optimized by choosing and blending salts.

Improvements in stability for aqueous electrolytes that do not reduce into a passivating film are not fully explained from first principles\cite{zheng_understanding_2018,yokoyama_origin_2018,wessells_investigations_2010}. Water-in-salt solutions are difficult to model with theories based on ideal solution behavior, and their properties are computationally expensive to calculate from first principles, with no guarantee of fidelity when compared to experiments. Thus, the rational design or computational screening of aqueous electrolytes may be challenging and time-consuming with very limited guarantee of success.
 
We reformulate aqueous electrolyte design as a black box optimization problem, where the electrolyte formulation is given as input and measured properties are output as optimization objectives. Our robotic electrolyte test-stand, named \textit{Otto}, mixes together aqueous electrolyte salts, pre-dissolved near saturation into feeder solutions, and measures two electrolyte objectives - ionic conductivity and electrochemical stability - along with temperature and pH. Electrochemical stability is tested with constant current holds at four current levels (111, 22, 5, then 1 mA/cm$^2$, first testing OER onset potentials then HER onset potentials) on two platinum wires with an Ag/AgCl reference electrode. As described previously \cite{whitacre_autonomous_2019}, we utilize a slope-extrapolation method between 22 and 5 mA/cm$^2$ to the zero-current axis to characterize electrolyte stability. This method will over-estimate electrolyte stability compared to longer measurements done at a lower currents (e.g. 50$\mu$ A/cm$^2$), but using this quantity during survey and optimization enables a 60 second measurement of electrolyte stability against HER and OER with consistent variance. Dosing, mixing, measurement, flushing, and washing steps meant that each experimental iteration took less than 25 minutes. In-depth details on the design, calibration, and performance of Otto and the fast electrochemical assessment are previously published\cite{whitacre_autonomous_2019}, but a detailed schematic of the test-stand mechanics and visualizations of the test are shown in the Extended Data Figs 1-4 and 8.

\begin{figure}
    \centering
    \includegraphics[width=0.75\textwidth]{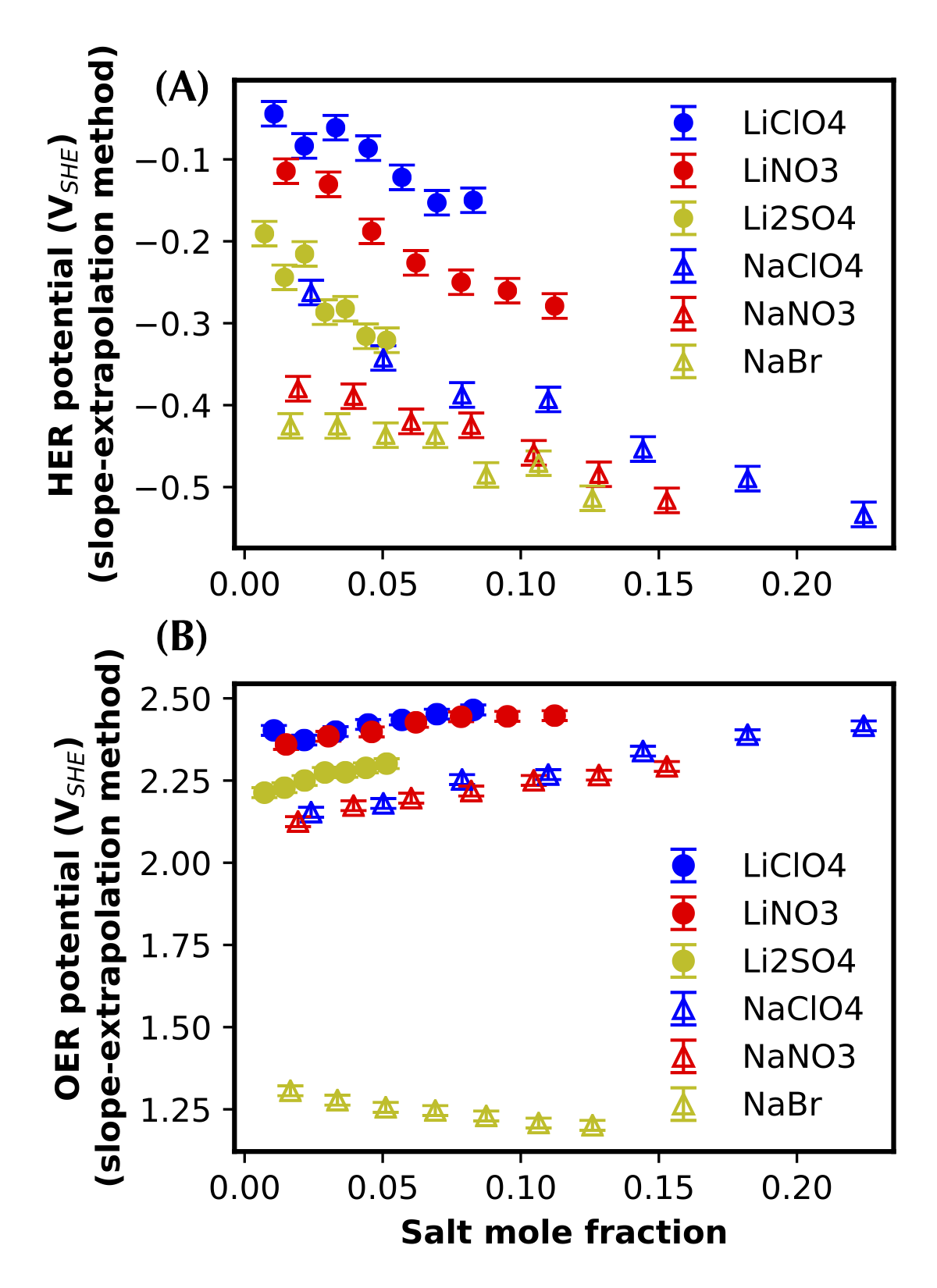}
    \caption{A manual grid survey of stability response to salt concentration for common aqueous electrolytes shows that a choice of salt greatly affects stability window. Figure (A) gives HER onset potential, and (B) for OER, using slope extrapolation method shifted to pH 0. This frames a design problem as choosing the best salt combination in water for widest stability window against HER and OER.}
    \label{fig:figure2}
\end{figure}

Otto first surveyed stability against HER and OER as a function of salt concentration on a manually-defined grid across common lithium and sodium electrolytes, without the machine-learning optimizer. Stability values are summarized using the slope-extrapolation method - results are shown as a function of mole fraction of salt in the electrolyte in Figure 2(A) and (B). All potentials reported in Figure 2 have been shifted to pH 0 with the Nernst equation, based on measured pH (see SI). Anions clearly perform differently in suppressing anode or cathode reactions on platinum. Outliers include \ce{Li2SO4} and \ce{NaBr} that diverge from nitrate and perchlorate salts, possibly due to unique stoichiometry or bromide oxidation respectively. Significant differences between nitrate and perchlorate salts are evident particularly on HER suppression. These results further motivate a design problem where salts are chosen and blended to discover a novel electrolyte with optimal stability window.

It is prohibitively time-consuming to exhaustively search mixtures of these anions for optimal formulations as the complexity of mixed electrolyte design spaces exhibits ``combinatorial explosion''. For example, Otto utilizes a testing volume of 7 mL. Exhaustively searching a space of 3-salt mixtures in 0.1 mL increments would require 62,000 evaluations, and a space of 4-salt mixtures would require 1.15 million evaluations. To make optimiztion over this design space practical, we connected Otto to Dragonfly, a Bayesian optimization software package developed by our team. Dragonfly harnesses a suite of acquisition strategies and evolutionary algorithms for scalable and robust treatment of black-box functions\cite{kandasamy_tuning_2019,PariaKP19}. Given only solubility constraints on mixtures, Dragonfly optimized for the electrochemical stability window - as measured by via the fast electrochemical assessment and summarized into a single number with slope-extrapolation method - over the design spaces of 1) mixtures of \ce{NaNO3}, \ce{NaClO4}, \ce{Na2SO4}, and \ce{NaBr} and 2) mixtures of \ce{LiNO3}, \ce{LiClO4}, and \ce{Li2SO4}. Dragonfly operated fully autonomously, running experiments with no human guidance. Results are illustrated in Figure 3(A-F). Concentrations of feeder solutions for each salt are given in Table 1; compositions of the best blended electrolytes discovered by Dragonfly are given in Table 2.

\begin{figure}
    \centering
    \includegraphics[scale=0.72]{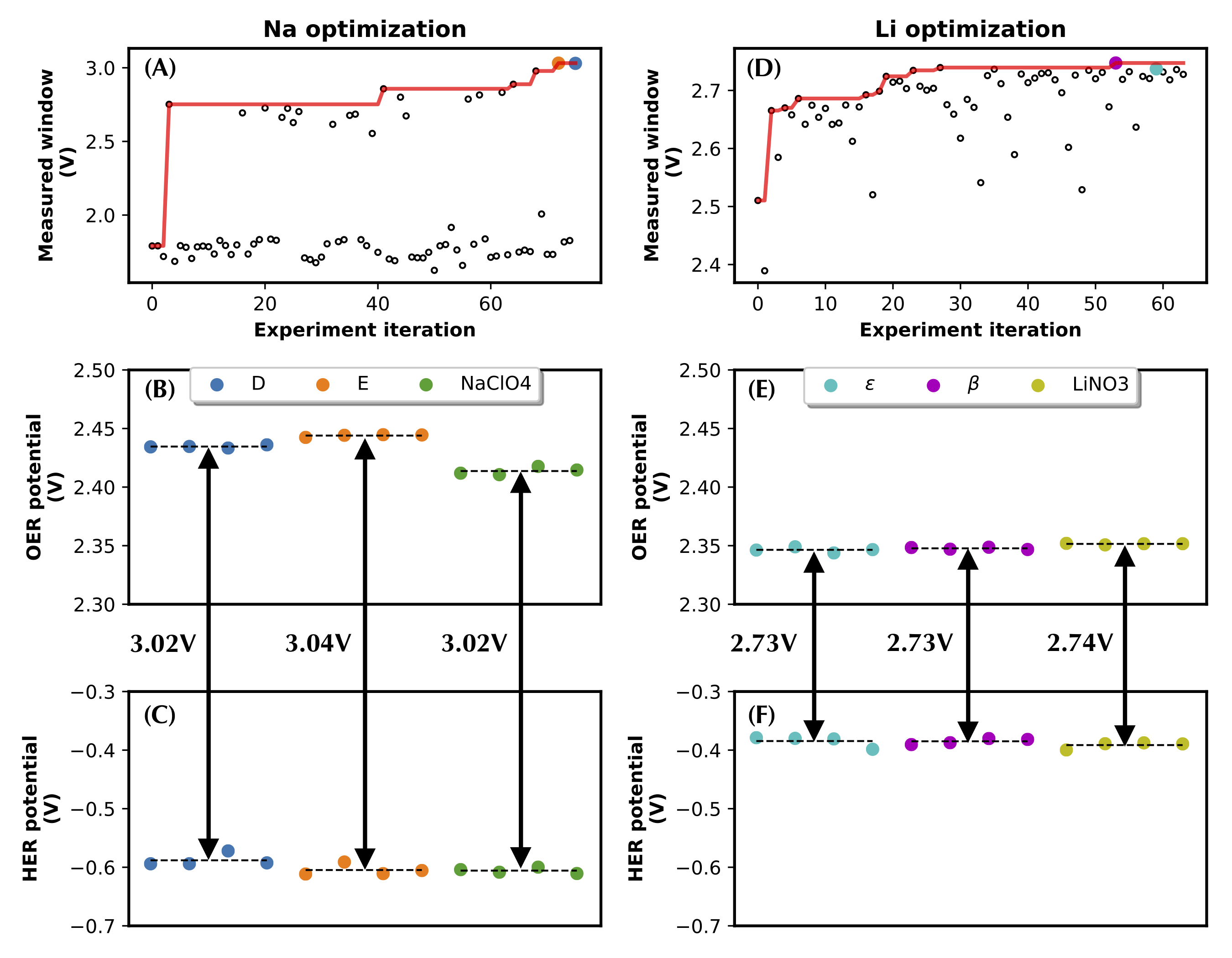}
    \caption{Results for autonomous optimization of sodium (left column) and lithium (right column) aqueous electrolytes. Top figures A and D shows Dragonfly's optimization routine, black circles are individual evaluations and the red line shows maximum found over iterations. Shown in color are the top blends found by Dragonfly, whose compositions are given in Table 1. For each top blend, 4 additional experiments were carried out against baselines of  \ce{NaClO4} and \ce{LiNO3} and potentials (via slope-extrapolation method) are reported in Figures B, C, E, and F. Blend E is the best performing sodium electrolyte and \ce{LiNO3} is the best lithium electrolyte.}
    \label{fig:figure3}
\end{figure}

\begin{table}
  \centering
    \begin{tabular}{l|l}
    Feeder Solution & Molality \\ \hline
     \ce{NaClO4} & 16.03 \\
    \ce{NaNO3} & 10.03 \\
     \ce{Na2SO4} & 1.5 \\
     \ce{NaBr} & 8 \\
     \ce{LiNO3} & 7.02 \\
     \ce{Li2SO4} & 3.01 \\
     \ce{LiClO4} & 5.01
    \end{tabular}
\caption{Concentrations of feeder solutions} \label{tab:table1}
\end{table}

\begin{table}
  \centering
    \begin{tabular}{l|l}
    Blend & Composition (mL of feeder solutions) \\ \hline
     D & 6.1 \ce{NaClO4}, 0.8 \ce{NaNO3}, 0.1 \ce{Na2SO4} \\
     E & 6.7 \ce{NaClO4}, 0.3 \ce{NaNO3} \\
     $\epsilon$ & 6.4 \ce{LiNO3}, 0.6 \ce{LiClO4}\\
     $\beta$ & 5.7 \ce{LiNO3}, 0.9 \ce{LiClO4}, 0.4 \ce{Li2SO4}
    \end{tabular}
\caption{Composition of electrolyte blends discovered; test volume was kept constant at 7mL.} \label{tab:table2}
\end{table}

The optimization curve over sodium electrolytes illustrated in Figure (A) is steep but flat in a middle portion, wherein Dragonfly was learning that any amount of \ce{NaBr} in the electrolyte significantly lowered cathode stability. This is an excellent illustration of the algorithm learning a nonlinear chemical response. Top blends from this optimization were run for an additional 4 experiments each against a pure \ce{NaClO4} feeder solution benchmark. The four measured potentials and their averages are reported in Figure 3 (B) and (C). Blend E showed an improved OER stability of 20 mV over to pure \ce{NaClO4}.

The optimization over lithium electrolytes is illustrated in Figure 3(D). Dragonfly initializes by randomly sampling the design space in the first five runs, which, in the lithium case, included a strong electrolyte. The three-dimensional design space is much smaller than the sodium design space. The lithium optimization converges much faster than the sodium optimization. The optimizer converged on two blends and pure \ce{LiNO3} feeder solution as three candidates with optimal stability windows - other high performing candidates were dilutions of \ce{LiNO3} and not tested. These electrolytes were run for an additional 4 experiments each; the measured potentials are shown in Figure 3(E) and (F). The concentrated \ce{LiNO3} electrolyte is the strongest performer tested by the optimizer and has been used extensively in literature\cite{li_rechargeable_1994,zheng_understanding_2018}.

\begin{figure}
    \centering
    \includegraphics[clip,width=.9\textwidth]{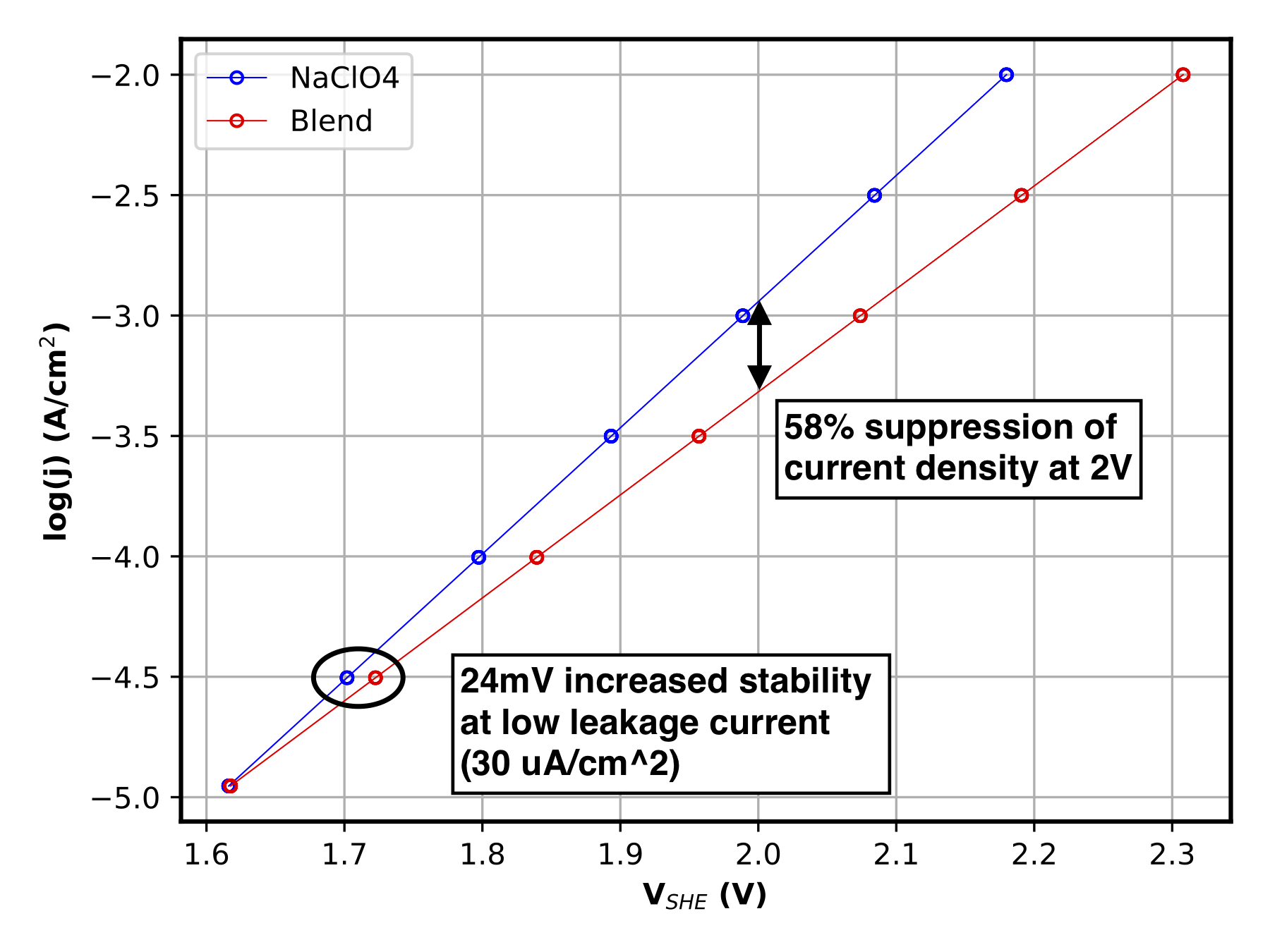}
    \caption{Results for 7 runs on Blend E against control \ce{NaClO4} suggest that the blend is better at suppressing OER than \ce{NaClO4}. The potential for an acceptable leakage current (30$\mu$A/cm$^2$) is 24 mV higher in the blend, and the blend illustrates significantly improved high-rate capability with a 58\% (-0.37 log units) suppression of current density at 2V compared to \ce{NaClO4}. The two electrolytes are close in pH (near 8.8); the potentials given are therefore not pH shifted.}
    \label{fig:figure4}
\end{figure}

Blend E and \ce{NaClO4} were run for a longer, detailed evaluation of OER stability in Otto, illustrated in Figure 4. Current density  was varied in half log-decade steps from $j=10^{-1}$ to $10^{-5} A/cm^2$. A Tafel equation was fit to the average of 7 sequential runs, ignoring high current steps $j=10^{-1}$ and $10^{-1.5} A/cm^2$. Otto has previously been used to replicate the Tafel slope of 1M KOH standard for OER posited in literature\cite{damjanovic_kinetics_1966} to within experimental error (Extended Data Fig.5). Full data figures and methods for this run are given in the SI and Extended Data Fig. 6 and 7. The results suggest that Blend E is more to stable to OER than \ce{NaClO4} - a high-performing state-of-the-art sodium electrolyte extensively evaluated in past work\cite{lee_toward_2019}. The potential at an acceptable OER leakage current\cite{wessells_investigations_2010} (30$\mu A/cm^2$) is 24 mV higher in the blend than in the \ce{NaClO4} feeder solution, suggesting a longer calendar life for a potential aqueous battery built with the blend. The blend also shows significantly improved high-rate capability with a 58\% suppression of OER current density at 2V compared to the \ce{NaClO4} feeder solution.

It is not a priori intuitive why a blend of two common sodium salts would better suppress OER compared to its pure counterparts - these two salts together may form a novel passivating film\cite{lee_toward_2019} or better suppress water activity\cite{zheng_understanding_2018}. Previous rationalization of OER suppression via the latter mechanism leveraged the Hofmeister series, a scale for relative interaction strength between a specific anion and water\cite{lee_toward_2019}. This may not apply to the potentially complex interactions in high-concentration, blended salts. Further theoretical and experimental analysis would be required to characterize the precise role nitrate anions play in the perchlorate electrolyte.

A machine-learning-guided robotic test-stand has explored a design space of previously reported aqueous electrolyte components, and discovered previously unknown, non-intuitive, but higher performing aqueous electrolyte. We believe this result serves as proof-of-principle that autonomous battery design can generate materials that a human designer may miss. More complex mixtures, whether aqueous or non-aqueous, and co-optimizing electrode and electrolyte can be tested without changing the principle of our approach.

\begin{methods}
\subsection{Materials}
Sodium perchlorate (CAS 7601-89-0, Anhydrous 98.0-102.0\%, ACS Grade) was purchased from VWR International; sodium nitrate (CAS 7631-99-4, ACS reagent, >99.0\%), sodium sulfate (CAS 7757-82-6, ACS reagent, Anhydrous, >99.0\%), sodium bromide (CAS 767-15-6, anhydrous >99\%), lithium perchlorate (CAS 7791-03-9, ACS reagent, >95.0\%), lithium nitrate (CAS 7790-69-4, Reagentplus), lithium sulfate (CAS 10377-48-7, Titration >98.5\%), and lithium bromide (CAS 7550-35-8, anhydrous, >99.\%) were purchased from Sigma Aldrich and were used without further purification.

\subsection{Material storage}
Bromide solids were stored and massed in a dry Argon atmosphere; sodium perchlorate solids were stored only as unopened containers; all other solids were stored in ambient lab atmosphere in parafilm sealed containers; massing of all non-bromide solids was conducted in ambient lab atmosphere. Electrolyte solutions were stored in sealed Fisher Scientific brand Kimble Kimax GL 45 containers at ambient lab conditions (22$^{\circ}$C $\pm$ 2$^{\circ}$C). Deionized water used for in test-stand dilution of solutions was stored exclusively in Fisher Scientific brand Kimble Kimax GL 45 containers.

\subsection{Preparation of solutions}
All solutions were mixed in ambient atmosphere, beakers were sealed off with parafilm once all solid was added to the deionized water. Solutions were mixed with VWR brand magnetic stir bars for a minimum of 30 min past the dissolution of the last visible solid. Temperature was regulated by setting hot plates to maximum of 30 $^{\circ}$C for endothermic solutions and through water baths at ambient temperature for exothermic solutions; all solutions were mixed for at least 30 minutes at ambient conditions before density measurements were performed.

\subsection{Experimental details}
Details on test-stand components for each measured property can be found in abundance in a previous publication\cite{whitacre_autonomous_2019}. As reported previously, each experimental iteration consisted of 3 separate Otto runs - a wash with deionized water, then an initial run with the requested mixture, then a second, ``production'' run that was reported as data. This procedure was shown to have highest fidelity against benchmark cases\cite{whitacre_autonomous_2019}. Testing volume was held constant at 7mL. Each machine learning optimization consisted of four days of experimentation, divided into daily runs of 10-15 iterations. Solutions were restocked as needed, and made according to above section in preparation. For detailed evaluations, see Extended Data Fig 6 and 7 for information. Platinum surfaces were cleaned in between the evaluation of \ce{NaClO4} and Blend E with 1200 grit wet paper, rinsed with isopropyl alcohol and deionized water.

\subsection{pH corrections}
The Nernst Equation shows that half-cell potential changes by 59mV for each pH unit change. Thus, to standardize for performance across the Pourbaix diagram, all reported potentials (unless otherwise noted) are shifted to pH zero, i.e.: $E_{reported}=E_{measured}+0.0591 pH_{measured}$

\subsection{Tafel equation}
The Tafel equation is derived from the Butler Volmer kinetic law: $j=j_0 \Big( exp\big[\frac{\alpha_a z e \eta}{k_b T}\big] - exp\big[\frac{-\alpha_c z e \eta}{k_b T}\big]\Big)$. Tafel approximation ignores the backward reaction and thus arrives at a functional form: $\eta = A\big[log(j)-log(j_0)\big]$. For Tafel plots presented in Figure 3, and Extended Data 5, 6, and 7, current density was converted to A/cm$^2$ and electrode potentials were reported on the standard hydrogen electrode scale.

\subsection{Combinatoric estimation of design spaces}
The design space for testing the mixtures described in this paper is constrained by Otto's testing volume (7mL) and can be described in equation form as: $x_1+x_2+...+x_d+x_{water}=7$ where each $x_i$ is the volume of feeder solution $i$ in the mixture, up to $d$ salts then diluting water (if present). If the design space were discretized into integer volumes, this becomes a form of the classic combinatorics problem in "stars and bars". For any pair of positive integers n and k, the number of k-tuples of non-negative integers whose sum is n is equal to: ${n+k-1}\choose{n}$. All electrolyte optimizations mentioned in paper occurred in discretizations of 0.1 mL. This maps directly to the above solution if the equation is multipled by 10, i.e.: $10 x_1 + 10 x_2 + ... + 10 x_d + 10 x_{water} = 70$. Thus, our analytic solution for the complexity of optiming across $d$ salts ($+1$ for $x_{water}$): ${70+(d+1)-1}\choose{70}$. ${73}\choose{70}$ = 62196. ${74}\choose{70}$ = 1150626.

\subsection{Data availability}
The supporting data for the included graphs within this paper, as well as other findings from this study, are available from the corresponding author upon reasonable request.

\subsection{Code availability}
The code for the plots presented in this paper is available from the corresponding author upon reasonable request.
\end{methods}

\bibliography{ML_otto}

\begin{thebibliography}{10}
\expandafter\ifx\csname url\endcsname\relax
  \def\url#1{\texttt{#1}}\fi
\expandafter\ifx\csname urlprefix\endcsname\relax\def\urlprefix{URL }\fi
\providecommand{\bibinfo}[2]{#2}
\providecommand{\eprint}[2][]{\url{#2}}

\bibitem{suo_water--salt_2015}
\bibinfo{author}{Suo, L.} \emph{et~al.}
\newblock \bibinfo{title}{``{Water}-in-salt'' electrolyte enables high-voltage
  aqueous lithium-ion chemistries}.
\newblock \emph{\bibinfo{journal}{Science}} \textbf{\bibinfo{volume}{350}},
  \bibinfo{pages}{938} (\bibinfo{year}{2015}).
\newblock
  \urlprefix\url{http://science.sciencemag.org/content/350/6263/938.abstract}.

\bibitem{zheng_understanding_2018}
\bibinfo{author}{Zheng, J.} \emph{et~al.}
\newblock \bibinfo{title}{Understanding {Thermodynamic} and {Kinetic}
  {Contributions} in {Expanding} the {Stability} {Window} of {Aqueous}
  {Electrolytes}}.
\newblock \emph{\bibinfo{journal}{Chem}} \textbf{\bibinfo{volume}{4}},
  \bibinfo{pages}{2872--2882} (\bibinfo{year}{2018}).
\newblock
  \urlprefix\url{https://linkinghub.elsevier.com/retrieve/pii/S2451929418304248}.

\bibitem{yokoyama_origin_2018}
\bibinfo{author}{Yokoyama, Y.}, \bibinfo{author}{Fukutsuka, T.},
  \bibinfo{author}{Miyazaki, K.} \& \bibinfo{author}{Abe, T.}
\newblock \bibinfo{title}{Origin of the {Electrochemical} {Stability} of
  {Aqueous} {Concentrated} {Electrolyte} {Solutions}}.
\newblock \emph{\bibinfo{journal}{J. Electrochem. Soc.}}
  \textbf{\bibinfo{volume}{165}}, \bibinfo{pages}{A3299--A3303}
  (\bibinfo{year}{2018}).
\newblock \urlprefix\url{http://jes.ecsdl.org/content/165/14/A3299}.

\bibitem{kandasamy_tuning_2019}
\bibinfo{author}{Kandasamy, K.} \emph{et~al.}
\newblock \bibinfo{title}{Tuning {Hyperparameters} without {Grad} {Students}:
  {Scalable} and {Robust} {Bayesian} {Optimisation} with {Dragonfly}}.
\newblock \emph{\bibinfo{journal}{arXiv:1903.06694 [cs, stat]}}
  (\bibinfo{year}{2019}).
\newblock \urlprefix\url{http://arxiv.org/abs/1903.06694}.
\newblock \bibinfo{note}{ArXiv: 1903.06694}.

\bibitem{Deng:2018aa}
\bibinfo{author}{Deng, J.}, \bibinfo{author}{Bae, C.},
  \bibinfo{author}{Marcicki, J.}, \bibinfo{author}{Masias, A.} \&
  \bibinfo{author}{Miller, T.}
\newblock \bibinfo{title}{Safety modelling and testing of lithium-ion batteries
  in electrified vehicles}.
\newblock \emph{\bibinfo{journal}{Nat. Energy}} \textbf{\bibinfo{volume}{3}},
  \bibinfo{pages}{261--266} (\bibinfo{year}{2018}).
\newblock \urlprefix\url{https://doi.org/10.1038/s41560-018-0122-3}.

\bibitem{Viswanathan2019}
\bibinfo{author}{Viswanathan, V.} \& \bibinfo{author}{Knapp, B.~M.}
\newblock \bibinfo{title}{Potential for electric aircraft}.
\newblock \emph{\bibinfo{journal}{Nat. Sustain.}} \textbf{\bibinfo{volume}{2}},
  \bibinfo{pages}{88--89} (\bibinfo{year}{2019}).
\newblock \urlprefix\url{https://doi.org/10.1038/s41893-019-0233-2}.

\bibitem{tabor_accelerating_2018}
\bibinfo{author}{Tabor, D.~P.} \emph{et~al.}
\newblock \bibinfo{title}{Accelerating the discovery of materials for clean
  energy in the era of smart automation}.
\newblock \emph{\bibinfo{journal}{Nat. Rev. Mater.}}
  \textbf{\bibinfo{volume}{3}}, \bibinfo{pages}{5--20} (\bibinfo{year}{2018}).
\newblock \urlprefix\url{https://www.nature.com/articles/s41578-018-0005-z}.

\bibitem{pmlr-v37-kandasamy15}
\bibinfo{author}{Kandasamy, K.}, \bibinfo{author}{Schneider, J.} \&
  \bibinfo{author}{Poczos, B.}
\newblock \bibinfo{title}{High dimensional bayesian optimisation and bandits
  via additive models}.
\newblock In \bibinfo{editor}{Bach, F.} \& \bibinfo{editor}{Blei, D.} (eds.)
  \emph{\bibinfo{booktitle}{Proceedings of the 32nd International Conference on
  Machine Learning}}, vol.~\bibinfo{volume}{37} of
  \emph{\bibinfo{series}{Proceedings of Machine Learning Research}},
  \bibinfo{pages}{295--304} (\bibinfo{publisher}{PMLR},
  \bibinfo{address}{Lille, France}, \bibinfo{year}{2015}).
\newblock \urlprefix\url{http://proceedings.mlr.press/v37/kandasamy15.html}.

\bibitem{pmlr-v97-kandasamy19a}
\bibinfo{author}{Kandasamy, K.} \emph{et~al.}
\newblock \bibinfo{title}{Myopic posterior sampling for adaptive goal oriented
  design of experiments}.
\newblock In \bibinfo{editor}{Chaudhuri, K.} \& \bibinfo{editor}{Salakhutdinov,
  R.} (eds.) \emph{\bibinfo{booktitle}{Proceedings of the 36th International
  Conference on Machine Learning}}, vol.~\bibinfo{volume}{97} of
  \emph{\bibinfo{series}{Proceedings of Machine Learning Research}},
  \bibinfo{pages}{3222--3232} (\bibinfo{publisher}{PMLR},
  \bibinfo{address}{Long Beach, California, USA}, \bibinfo{year}{2019}).
\newblock \urlprefix\url{http://proceedings.mlr.press/v97/kandasamy19a.html}.

\bibitem{PariaKP19}
\bibinfo{author}{Paria, B.}, \bibinfo{author}{Kandasamy, K.} \&
  \bibinfo{author}{P{\'{o}}czos, B.}
\newblock \bibinfo{title}{A flexible framework for multi-objective bayesian
  optimization using random scalarizations}.
\newblock In \emph{\bibinfo{booktitle}{Proceedings of the Thirty-Fifth
  Conference on Uncertainty in Artificial Intelligence, {UAI} 2019, Tel Aviv,
  Israel, July 22-25, 2019}}, \bibinfo{pages}{267} (\bibinfo{year}{2019}).
\newblock \urlprefix\url{http://auai.org/uai2019/proceedings/papers/267.pdf}.

\bibitem{HernandezLobato16}
\bibinfo{author}{Hern\'andez-Lobato, D.}, \bibinfo{author}{Hern\'andez-Lobato,
  J.~M.}, \bibinfo{author}{Shah, A.} \& \bibinfo{author}{Adams, R.~P.}
\newblock \bibinfo{title}{Predictive entropy search for multi-objective
  bayesian optimization}.
\newblock In \emph{\bibinfo{booktitle}{ICML}}, \bibinfo{pages}{1492--1501}
  (\bibinfo{year}{2016}).
\newblock
  \urlprefix\url{http://proceedings.mlr.press/v48/hernandez-lobatoa16.html}.

\bibitem{kusne_fly_2014}
\bibinfo{author}{Kusne, A.~G.} \emph{et~al.}
\newblock \bibinfo{title}{On-the-fly machine-learning for high-throughput
  experiments: search for rare-earth-free permanent magnets}.
\newblock \emph{\bibinfo{journal}{Sci. Rep.}} \textbf{\bibinfo{volume}{4}},
  \bibinfo{pages}{6367} (\bibinfo{year}{2014}).
\newblock \urlprefix\url{https://www.nature.com/articles/srep06367}.

\bibitem{zunger_inverse_2018}
\bibinfo{author}{Zunger, A.}
\newblock \bibinfo{title}{Inverse design in search of materials with target
  functionalities}.
\newblock \emph{\bibinfo{journal}{Nat. Rev. Chem.}}
  \textbf{\bibinfo{volume}{2}}, \bibinfo{pages}{1--16} (\bibinfo{year}{2018}).
\newblock \urlprefix\url{https://www.nature.com/articles/s41570-018-0121}.

\bibitem{granda_controlling_2018}
\bibinfo{author}{Granda, J.~M.}, \bibinfo{author}{Donina, L.},
  \bibinfo{author}{Dragone, V.}, \bibinfo{author}{Long, D.-L.} \&
  \bibinfo{author}{Cronin, L.}
\newblock \bibinfo{title}{Controlling an organic synthesis robot with machine
  learning to search for new reactivity}.
\newblock \emph{\bibinfo{journal}{Nature}} \textbf{\bibinfo{volume}{559}},
  \bibinfo{pages}{377--381} (\bibinfo{year}{2018}).
\newblock \urlprefix\url{https://www.nature.com/articles/s41586-018-0307-8}.

\bibitem{bhowmik_perspective_2019}
\bibinfo{author}{Bhowmik, A.} \emph{et~al.}
\newblock \bibinfo{title}{A perspective on inverse design of battery
  interphases using multi-scale modelling, experiments and generative deep
  learning}.
\newblock \emph{\bibinfo{journal}{Energy Storage Mater.}}
  \textbf{\bibinfo{volume}{21}}, \bibinfo{pages}{446--456}
  (\bibinfo{year}{2019}).
\newblock
  \urlprefix\url{http://www.sciencedirect.com/science/article/pii/S2405829719302193}.

\bibitem{bai_accelerated_2019}
\bibinfo{author}{Bai, Y.} \emph{et~al.}
\newblock \bibinfo{title}{Accelerated {Discovery} of {Organic} {Polymer}
  {Photocatalysts} for {Hydrogen} {Evolution} from {Water} through the
  {Integration} of {Experiment} and {Theory}}.
\newblock \emph{\bibinfo{journal}{J. Am. Chem. Soc.}}
  \textbf{\bibinfo{volume}{141}}, \bibinfo{pages}{9063--9071}
  (\bibinfo{year}{2019}).
\newblock \urlprefix\url{https://doi.org/10.1021/jacs.9b03591}.

\bibitem{sun_accelerated_2019}
\bibinfo{author}{Sun, S.} \emph{et~al.}
\newblock \bibinfo{title}{Accelerated {Development} of {Perovskite}-{Inspired}
  {Materials} via {High}-{Throughput} {Synthesis} and {Machine}-{Learning}
  {Diagnosis}}.
\newblock \emph{\bibinfo{journal}{Joule}} \textbf{\bibinfo{volume}{3}},
  \bibinfo{pages}{1437--1451} (\bibinfo{year}{2019}).
\newblock
  \urlprefix\url{http://www.sciencedirect.com/science/article/pii/S2542435119302570}.

\bibitem{nikolaev_discovery_2014}
\bibinfo{author}{Nikolaev, P.}, \bibinfo{author}{Hooper, D.},
  \bibinfo{author}{Perea-L{\'o}pez, N.}, \bibinfo{author}{Terrones, M.} \&
  \bibinfo{author}{Maruyama, B.}
\newblock \bibinfo{title}{Discovery of {Wall}-{Selective} {Carbon} {Nanotube}
  {Growth} {Conditions} via {Automated} {Experimentation}}.
\newblock \emph{\bibinfo{journal}{ACS Nano}} \textbf{\bibinfo{volume}{8}},
  \bibinfo{pages}{10214--10222} (\bibinfo{year}{2014}).
\newblock \urlprefix\url{https://doi.org/10.1021/nn503347a}.

\bibitem{li_rapid_2017}
\bibinfo{author}{Li, C.} \emph{et~al.}
\newblock \bibinfo{title}{Rapid {Bayesian} optimisation for synthesis of short
  polymer fiber materials}.
\newblock \emph{\bibinfo{journal}{Sci. Rep.}} \textbf{\bibinfo{volume}{7}},
  \bibinfo{pages}{1--10} (\bibinfo{year}{2017}).
\newblock \urlprefix\url{https://www.nature.com/articles/s41598-017-05723-0}.

\bibitem{bessa2019}
\bibinfo{author}{Bessa, M.~A.}, \bibinfo{author}{Glowacki, P.} \&
  \bibinfo{author}{Houlder, M.}
\newblock \bibinfo{title}{Bayesian machine learning in metamaterial design:
  Fragile becomes supercompressible}.
\newblock \emph{\bibinfo{journal}{Adv. Mater.}} \textbf{\bibinfo{volume}{0}},
  \bibinfo{pages}{1904845} (\bibinfo{year}{2019}).
\newblock
  \urlprefix\url{https://onlinelibrary.wiley.com/doi/abs/10.1002/adma.201904845}.
\newblock
  \eprint{https://onlinelibrary.wiley.com/doi/pdf/10.1002/adma.201904845}.

\bibitem{macleod_self-driving_2019}
\bibinfo{author}{MacLeod, B.~P.} \emph{et~al.}
\newblock \bibinfo{title}{Self-driving laboratory for accelerated discovery of
  thin-film materials}.
\newblock \emph{\bibinfo{journal}{arXiv:1906.05398 [cond-mat,
  physics:physics]}}  (\bibinfo{year}{2019}).
\newblock \urlprefix\url{http://arxiv.org/abs/1906.05398}.
\newblock \bibinfo{note}{ArXiv: 1906.05398}.

\bibitem{whitacre_autonomous_2019}
\bibinfo{author}{Whitacre, J.}, \bibinfo{author}{Mitchell, J.},
  \bibinfo{author}{Dave, A.}, \bibinfo{author}{Burke, S.} \&
  \bibinfo{author}{Viswanathan, V.}
\newblock \bibinfo{title}{An autonomous electrochemical test stand for machine
  learning informed electrolyte optimization}  (\bibinfo{year}{2019}).
\newblock
  \urlprefix\url{https://chemrxiv.org/articles/An_Autonomous_Electrochemical_Test_stand_for_Machine_Learning_Informed_Electrolyte_Optimization/9971741}.

\bibitem{dave_benchmarking_2020}
\bibinfo{author}{Dave, A.}, \bibinfo{author}{Gering, K.~L.},
  \bibinfo{author}{Mitchell, J.~M.}, \bibinfo{author}{Whitacre, J.} \&
  \bibinfo{author}{Viswanathan, V.}
\newblock \bibinfo{title}{Benchmarking {Conductivity} {Predictions} of the
  {Advanced} {Electrolyte} {Model} ({AEM}) for {Aqueous} {Systems}}.
\newblock \emph{\bibinfo{journal}{J. Electrochem. Soc.}}
  \textbf{\bibinfo{volume}{167}}, \bibinfo{pages}{013514}
  (\bibinfo{year}{2020}).
\newblock \urlprefix\url{http://jes.ecsdl.org/content/167/1/013514}.

\bibitem{lee_toward_2019}
\bibinfo{author}{Lee, M.~H.} \emph{et~al.}
\newblock \bibinfo{title}{Toward a low-cost high-voltage sodium aqueous
  rechargeable battery}.
\newblock \emph{\bibinfo{journal}{Mater. Today}}  (\bibinfo{year}{2019}).
\newblock
  \urlprefix\url{http://www.sciencedirect.com/science/article/pii/S1369702118312604}.

\bibitem{luo_raising_2010}
\bibinfo{author}{Luo, J.-Y.}, \bibinfo{author}{Cui, W.-J.},
  \bibinfo{author}{He, P.} \& \bibinfo{author}{Xia, Y.-Y.}
\newblock \bibinfo{title}{Raising the cycling stability of aqueous lithium-ion
  batteries by eliminating oxygen in the electrolyte}.
\newblock \emph{\bibinfo{journal}{Nat. Chem.}} \textbf{\bibinfo{volume}{2}},
  \bibinfo{pages}{760--765} (\bibinfo{year}{2010}).
\newblock \urlprefix\url{http://www.nature.com/articles/nchem.763}.

\bibitem{li_towards_2013}
\bibinfo{author}{Li, Z.}, \bibinfo{author}{Young, D.}, \bibinfo{author}{Xiang,
  K.}, \bibinfo{author}{Carter, W.~C.} \& \bibinfo{author}{Chiang, Y.-M.}
\newblock \bibinfo{title}{Towards {High} {Power} {High} {Energy} {Aqueous}
  {Sodium}-{Ion} {Batteries}: {The} {NaTi}2({PO}4)3/{Na}0.44mno2 {System}}.
\newblock \emph{\bibinfo{journal}{Adv. Energy Mater.}}
  \textbf{\bibinfo{volume}{3}}, \bibinfo{pages}{290--294}
  (\bibinfo{year}{2013}).
\newblock
  \urlprefix\url{https://onlinelibrary.wiley.com/doi/abs/10.1002/aenm.201200598}.

\bibitem{whitacre_polyionic_2015}
\bibinfo{author}{Whitacre, J.~F.} \emph{et~al.}
\newblock \bibinfo{title}{A {Polyionic}, {Large}-{Format} {Energy} {Storage}
  {Device} {Using} an {Aqueous} {Electrolyte} and {Thick}-{Format} {Composite}
  {NaTi}2({PO}4)3/{Activated} {Carbon} {Negative} {Electrodes}}.
\newblock \emph{\bibinfo{journal}{Energy Technol.}}
  \textbf{\bibinfo{volume}{3}}, \bibinfo{pages}{20--31} (\bibinfo{year}{2015}).
\newblock
  \urlprefix\url{https://onlinelibrary.wiley.com/doi/abs/10.1002/ente.201402127}.

\bibitem{wu_relating_2015}
\bibinfo{author}{Wu, W.}, \bibinfo{author}{Shabhag, S.},
  \bibinfo{author}{Chang, J.}, \bibinfo{author}{Rutt, A.} \&
  \bibinfo{author}{Whitacre, J.~F.}
\newblock \bibinfo{title}{Relating {Electrolyte} {Concentration} to
  {Performance} and {Stability} for {NaTi}2({PO}4)3/{Na}0.44mno2 {Aqueous}
  {Sodium}-{Ion} {Batteries}}.
\newblock \emph{\bibinfo{journal}{J. Electrochem. Soc.}}
  \textbf{\bibinfo{volume}{162}}, \bibinfo{pages}{A803--A808}
  (\bibinfo{year}{2015}).
\newblock \urlprefix\url{http://jes.ecsdl.org/content/162/6/A803}.

\bibitem{li_rechargeable_1994}
\bibinfo{author}{Li, W.}, \bibinfo{author}{Dahn, J.~R.} \&
  \bibinfo{author}{Wainwright, D.~S.}
\newblock \bibinfo{title}{Rechargeable {Lithium} {Batteries} with {Aqueous}
  {Electrolytes}}.
\newblock \emph{\bibinfo{journal}{Science}} \textbf{\bibinfo{volume}{264}},
  \bibinfo{pages}{1115--1118} (\bibinfo{year}{1994}).
\newblock
  \urlprefix\url{http://www.sciencemag.org/cgi/doi/10.1126/science.264.5162.1115}.

\bibitem{luo_aqueous_2007}
\bibinfo{author}{Luo, J.-Y.} \& \bibinfo{author}{Xia, Y.-Y.}
\newblock \bibinfo{title}{Aqueous {Lithium}-ion {Battery}
  {LiTi}2({PO}4)3/{LiMn}2o4 with {High} {Power} and {Energy} {Densities} as
  well as {Superior} {Cycling} {Stability}}.
\newblock \emph{\bibinfo{journal}{Adv. Funct. Mater.}}
  \textbf{\bibinfo{volume}{17}}, \bibinfo{pages}{3877--3884}
  (\bibinfo{year}{2007}).
\newblock
  \urlprefix\url{https://onlinelibrary.wiley.com/doi/abs/10.1002/adfm.200700638}.

\bibitem{weber_long_2019}
\bibinfo{author}{Weber, R.} \emph{et~al.}
\newblock \bibinfo{title}{Long cycle life and dendrite-free lithium morphology
  in anode-free lithium pouch cells enabled by a dual-salt liquid electrolyte}.
\newblock \emph{\bibinfo{journal}{Nat. Energy}} \textbf{\bibinfo{volume}{4}},
  \bibinfo{pages}{683--689} (\bibinfo{year}{2019}).
\newblock \urlprefix\url{http://www.nature.com/articles/s41560-019-0428-9}.

\bibitem{suo_advanced_2016}
\bibinfo{author}{Suo, L.} \emph{et~al.}
\newblock \bibinfo{title}{Advanced {High}-{Voltage} {Aqueous} {Lithium}-{Ion}
  {Battery} {Enabled} by ``{Water}-in-{Bisalt}'' {Electrolyte}}.
\newblock \emph{\bibinfo{journal}{Angew. Chem.}} \textbf{\bibinfo{volume}{55}},
  \bibinfo{pages}{7136--7141} (\bibinfo{year}{2016}).
\newblock
  \urlprefix\url{https://onlinelibrary.wiley.com/doi/abs/10.1002/anie.201602397}.

\bibitem{wessells_investigations_2010}
\bibinfo{author}{Wessells, C.}, \bibinfo{author}{Ruffo, R.},
  \bibinfo{author}{Huggins, R.~A.} \& \bibinfo{author}{Cui, Y.}
\newblock \bibinfo{title}{Investigations of the {Electrochemical} {Stability}
  of {Aqueous} {Electrolytes} for {Lithium} {Battery} {Applications}}.
\newblock \emph{\bibinfo{journal}{Electrochem. Solid-State Lett.}}
  \textbf{\bibinfo{volume}{13}}, \bibinfo{pages}{A59} (\bibinfo{year}{2010}).
\newblock \urlprefix\url{http://esl.ecsdl.org/cgi/doi/10.1149/1.3329652}.

\bibitem{damjanovic_kinetics_1966}
\bibinfo{author}{Damjanovic, A.}, \bibinfo{author}{Dey, A.} \&
  \bibinfo{author}{Bockris, J.}
\newblock \bibinfo{title}{Kinetics of oxygen evolution and dissolution on
  platinum electrodes}.
\newblock \emph{\bibinfo{journal}{Electrochim. Acta}}
  \textbf{\bibinfo{volume}{11}}, \bibinfo{pages}{791--814}
  (\bibinfo{year}{1966}).
\newblock
  \urlprefix\url{https://linkinghub.elsevier.com/retrieve/pii/0013468666870561}.

\end{thebibliography}

\begin{addendum}
\item This work was supported by Toyota Research Institute through the Accelerated Materials Design and Discovery program.  The authors acknowledge insightful discussions with Brian Storey, Abraham Anapolsky, Linda Hung and Chirranjeevi Gopal from the Toyota Research Institute.
\item[Contributions] J. W., V. V. and Barnabas P. conceived the project. J. M. designed, machined, and assembled the test-stand and wrote Labview control software. J.M. designed the fast electrochemical assessment. A. D. designed Python software and data layer and web-server interface. S. B. prepared all feeder solutions and stocked the test-stand. A. D. managed experiments (data input and output for test-stand both in manually defined and machine-learning operated modes) and analyzed results. K. K. and Biswajit P. wrote Dragonfly, consulted on its applicability to this problem, and implemented required features for running Otto from Dragonfly.  A. D. and V. V. wrote the paper with input from all the authors. A. D. and J. M. produced all the figures.
\item[Corresponding Authors] Correspondence and requests for materials should be addressed to Jay Whitacre, ~(email: whitacre@andrew.cmu.edu) and Venkat Viswanathan, ~(email: venkvis@cmu.edu).
\item[Competing Interests] The authors declare that they have no competing financial interests.

\end{addendum}



\section*{Supplementary material}

# Extended Data

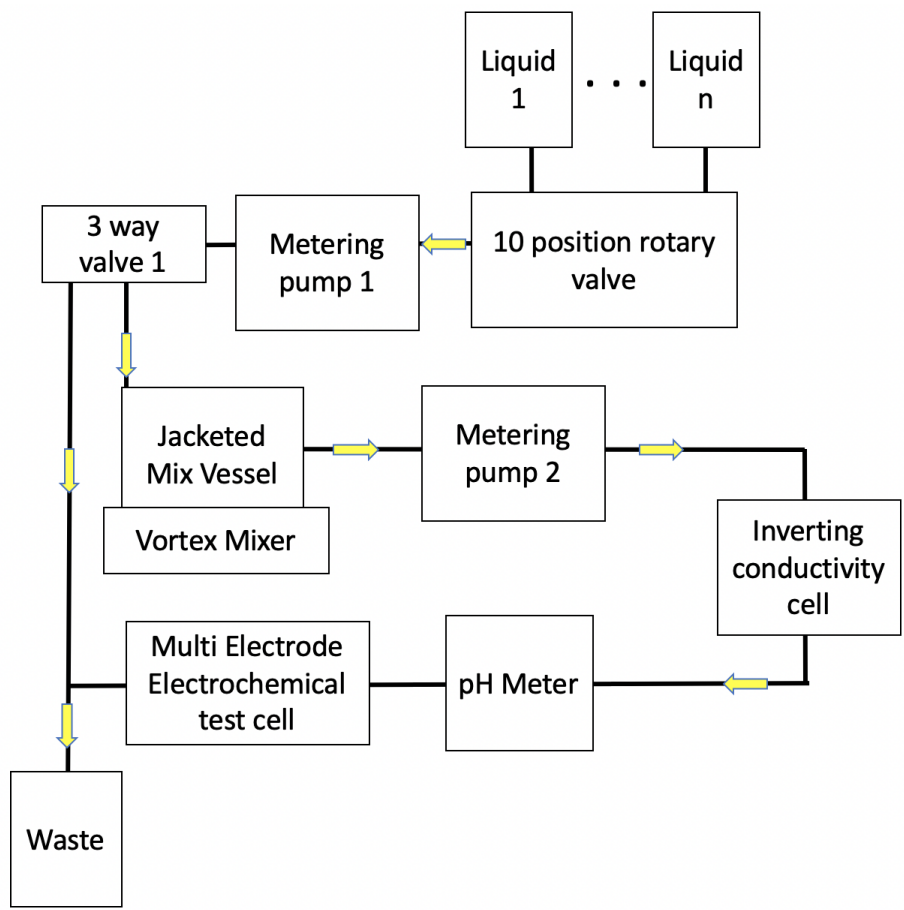

**Figure 1:** Flow schematic of Otto - rotary valve manages feeder solutions with near-saturation, single salts. Mixing occurs in a jacketed mix vessel, where pH metering occurs. Conductivity readings are taken before electrochemical testing.

| Blend | Composition (by vol% of feeder solutions) |
|---|---|
| D | $NaClO_4$:$NaNO_3$:$Na_2SO_4$ 87:12:1 |
| E | $NaClO_4$:$NaNO_3$ 96:4 |
| $\alpha$ | $LiNO_3$:$LiClO_4$:$Li_2SO_4$ 92:7:1 |
| $\delta$ | $LiNO_3$:$LiClO_4$:$Li_2SO_4$ 90:6:4 |

**Table 1:** Composition of electrolyte blends discovered

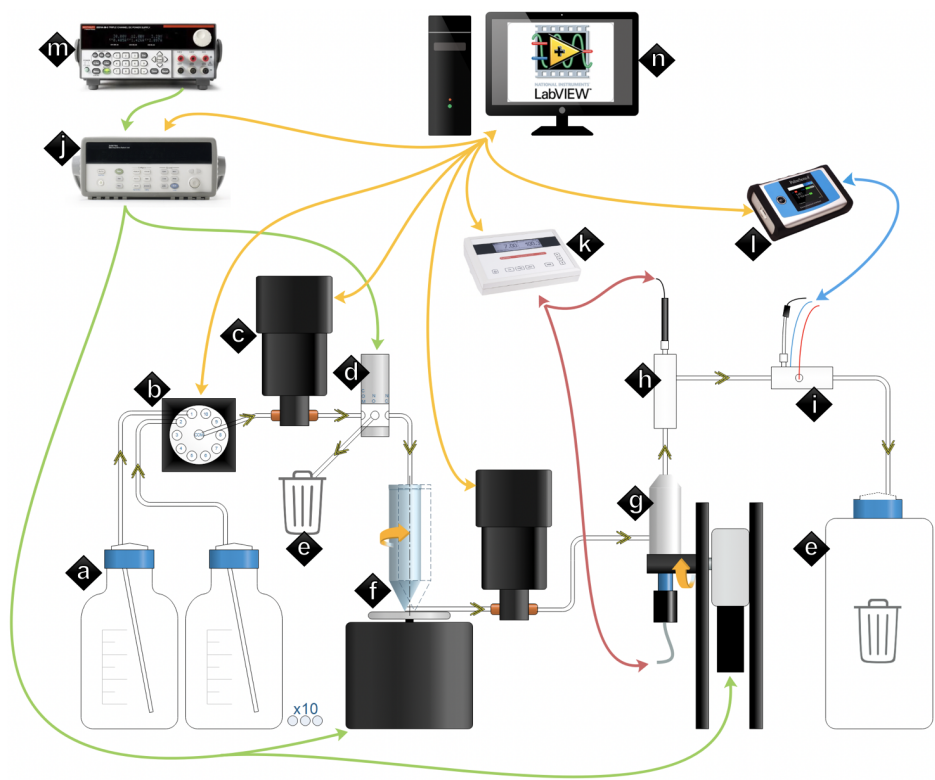

**Figure 2:** A: feeder solutions; B: VICI rotary valve; C: positive displacement pumps made by Fluid Metering; D: three-way valve; E: waste vessel; F: jacketed rotary mixer; G: inverting conductivity chamber with Consort probe (K); H: pH meter also connected to Consort; I: electrochemical testing chamber with Palmsens potentiostat. J: multiplexer for all switches and triggers; M: power supply.

3

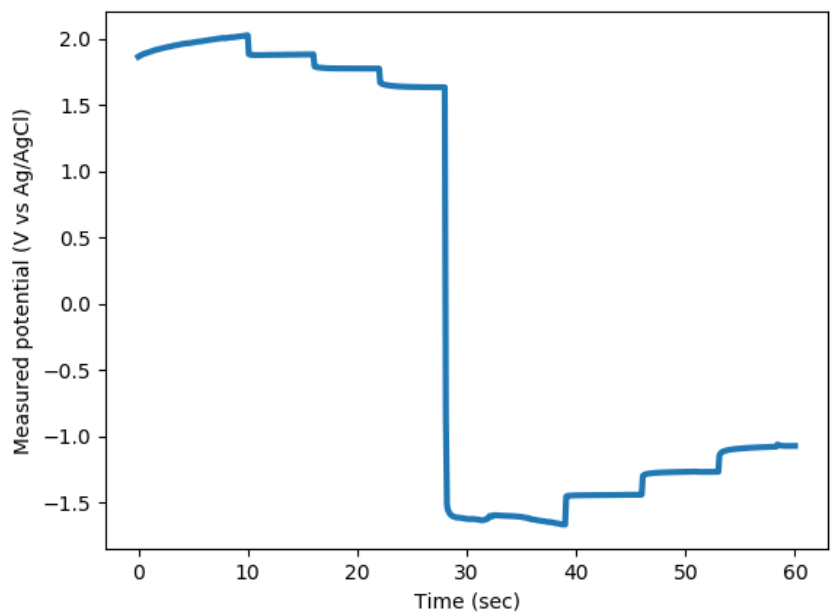

**Figure 3:** Raw visualization of staircase potentiometry over time - measured potential over time. Each step corresponds to a different current density, potential is measured by the Palmsens probes attached to platinum electrodes with respect to an Ag/AgCl reference. The electrolyte tested here is 7mL of 16 molal NaClO4 dissolved in water.

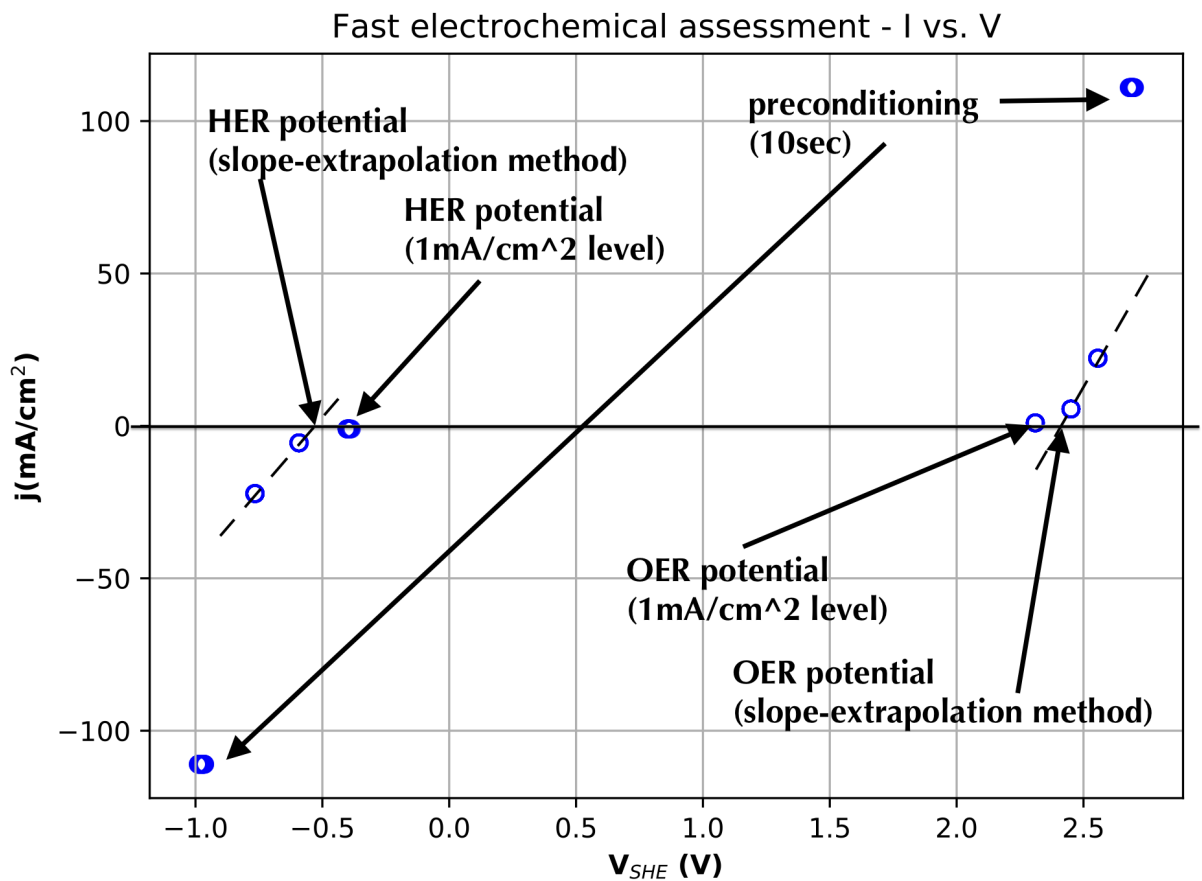

**Figure 4:** Derived quantities from staircase potentiometry visualized. Quantities are calculated using the average of the last 2 seconds of each current step (to sample the equilibrated data). The surface is preconditioned with 10seconds of 111 mA/cm$^2$ current on each side. The order of the test is positive (i.e. anodic, OER) currents at 111, 20, 5, then 1 mA/cm$^2$, then the same currents as negative (i.e. cathodic, HER), with non-preconditioning steps held for 6-7 seconds. The slope-extrapolation method has consistent performance[1] across a range of electrolytes, and thus is used during grid survey and ML-guided optimization. Afterwards, the best blends were subjected to a detailed evaluation at many current steps (much lower than the lowest in this fast assessment). The electrolyte tested in this figure is 7mL of 16 molal NaClO4 dissolved in water, pH shifted to 0 and reported against SHE (converted from Ag/AgCl).

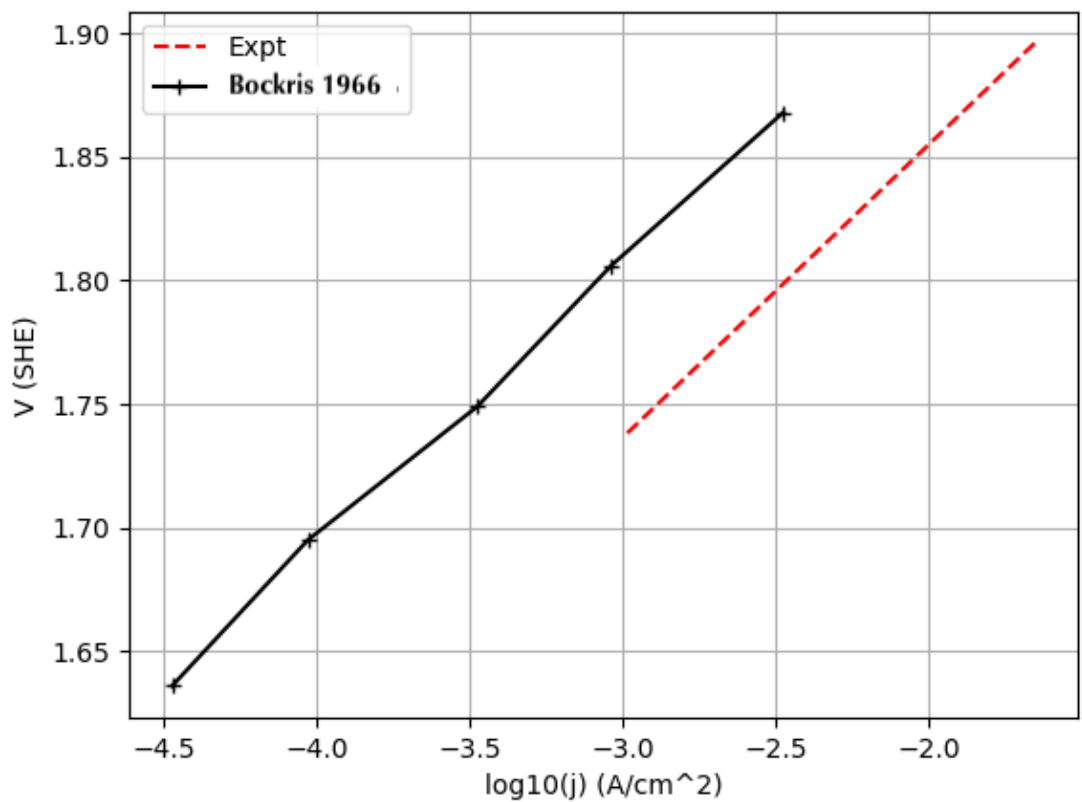

**Figure 5:** Tafel extrapolation of Otto's staircase potentiometry / fast assessment experiment was benchmarked against literature results for 1M KOH[2]. 25 experiments were run and Tafel lines were fit between voltage (against standard hydrogen electrode) and log current density. The results reproduce the Tafel slope reported in the 1966 paper within their reported experimental uncertainty. Potential difference of 50mV could be due to electrode surfaces or geometry differences.

6

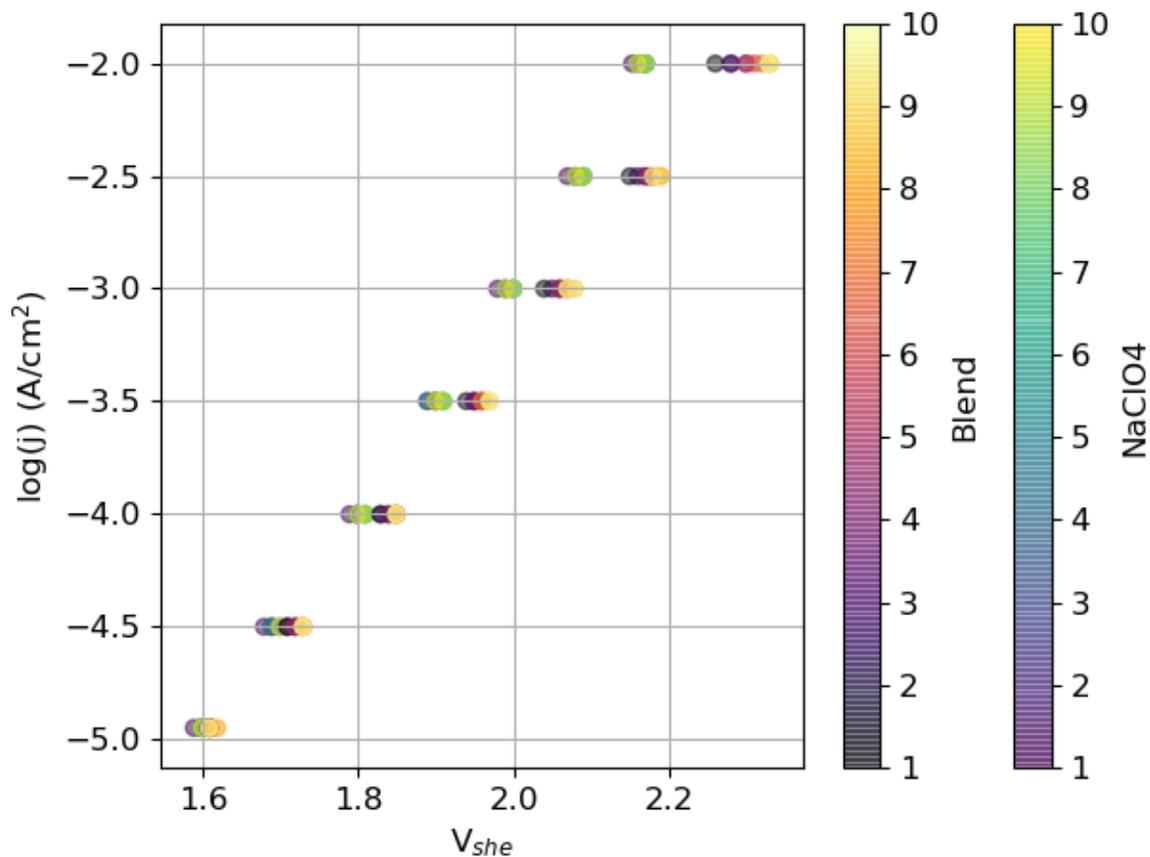

**Figure 6:** A detailed evaluation of Blend E and NaClO$_4$ was undertaken by running half-log-decade steps of current density from $10^{-1}$ to $10^{-5}$ in $A/cm^2$. Only current densities from -2 and -5 were considered for the fit. 10 sequential runs were completed (run order is illustrated in the two colorbars), recirculating electrolyte; based on the above figure, the first three runs for each electrolyte were ignored as results indicated the conditioning/formation of platinum oxide film on the electrodes. Steps $log(j) = -2$ to $-3.5$ were held for 30 seconds each, with the last 5 seconds averaged to generate the data point. Step $log(j) = -4$ was held for 60 seconds, $-4.5$ for 90 seconds, and $-5$ for 120 seconds, with the last 5 seconds of each averaged to get each point plotted.

7

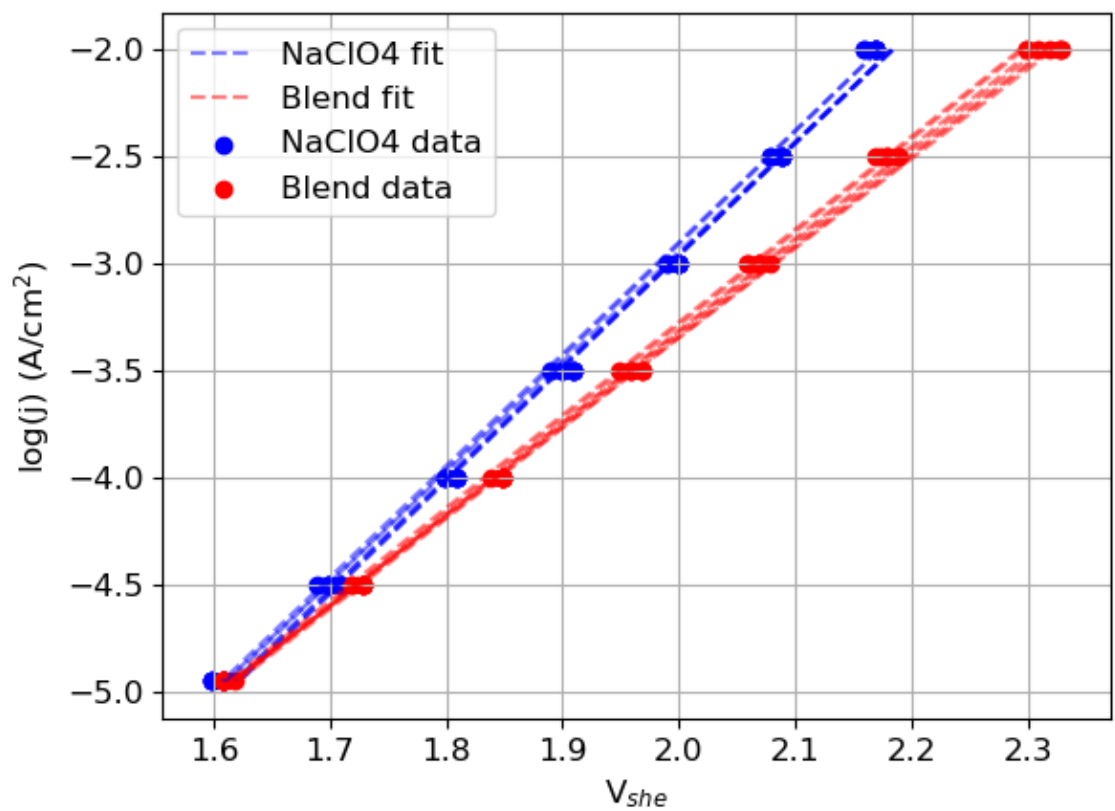

**Figure 7:** Each of the 7 retained runs were fit by a separate linear Tafel equation. These parameters were averaged, resulting in the single curve illustrated in the main paper Figure 3.

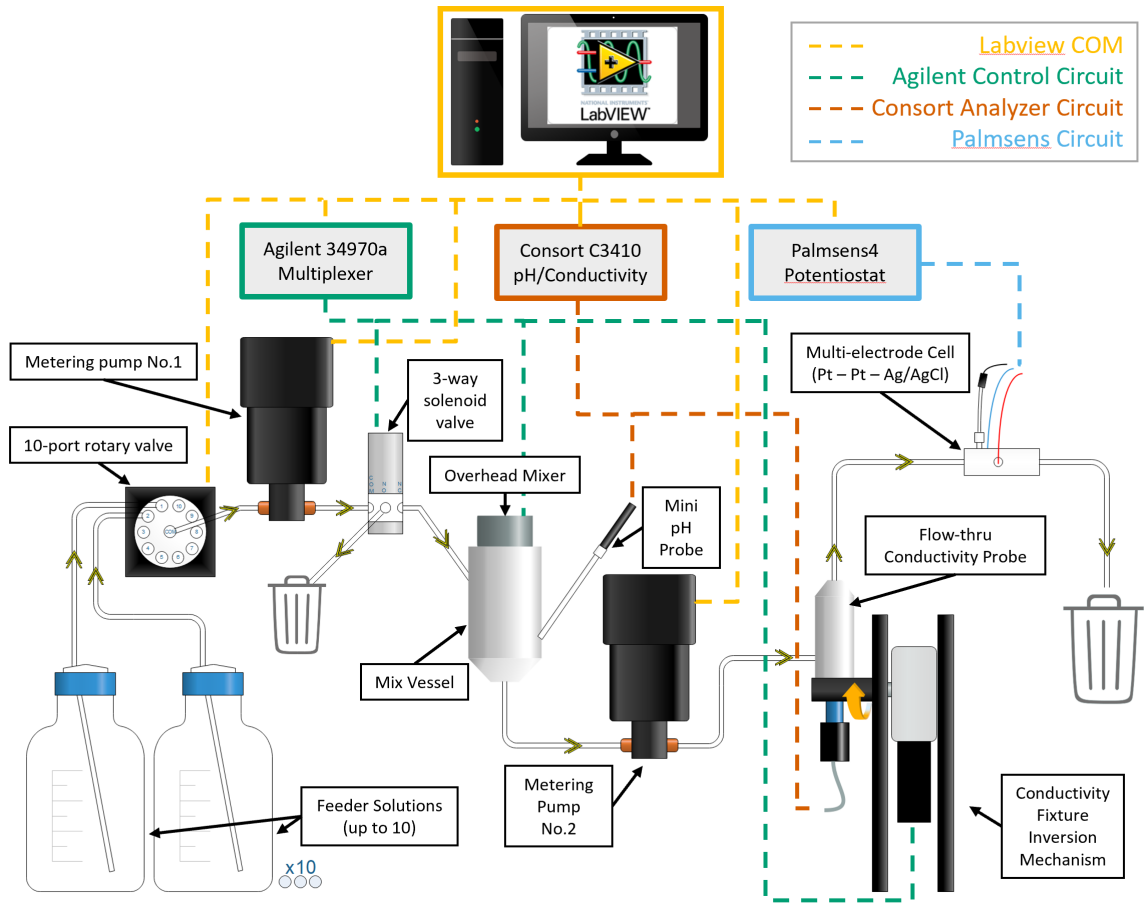

**Figure 8:** Detailed schematic of test-stand. Measurements are taken via Consort meter, Palmsens programmable potentiostat, and switching is handled through Agilent multiplexer.

# References

1. Whitacre, J., Mitchell, J., Dave, A., Burke, S. & Viswanathan, V. An autonomous electrochemical test stand for machine learning informed electrolyte optimization. *ChemrXiv, DOI: 10.26434/chemrxiv.9971741.v1* (2019).

2. Damjanovic, A., Dey, A. & Bockris, J. Kinetics of oxygen evolution and dissolution on platinum electrodes. *Electrochim. Acta* **11**, 791–814 (1966). URL `https://linkinghub.elsevier.com/retrieve/pii/0013468666870561`.



\end{document}